\documentstyle[12pt]{article}

\font\twlgot =eufm10 scaled \magstep1
\font\egtgot =eufm8
\font\sevgot =eufm7

\font\twlmsb =msbm10 scaled \magstep1
\font\egtmsb =msbm8
\font\sevmsb =msbm7

\newfam\gotfam
\def\pgot{\fam\gotfam\twlgot}
\textfont\gotfam\twlgot
\scriptfont\gotfam\egtgot
\scriptscriptfont\gotfam\sevgot
\def\got{\protect\pgot}

\newfam\msbfam
\textfont\msbfam\twlmsb
\scriptfont\msbfam\egtmsb
\scriptscriptfont\msbfam\sevmsb
\def\Bbb{\protect\pBbb}
\def\pBbb{\relax\ifmmode\expandafter\Bb\else\typeout{You cann't use
Bbb in text mode}\fi}
\def\Bb #1{{\fam\msbfam\relax#1}}

\def\op#1{\mathop{{\it\fam0} #1}\limits}

\newcommand{\Ker}{{\rm Ker\,}}
\newcommand{\im}{{\rm Im\, }}
\newcommand{\nm}[1]{\mid {#1}\mid}
\newcommand{\pr}{{\rm pr}}
\newcommand{\beq}{\begin{equation}}
\newcommand{\eeq}{\end{equation}}
\newcommand{\ben}{\begin{eqnarray}}
\newcommand{\een}{\end{eqnarray}}
\newcommand{\be}{\begin{eqnarray*}}
\newcommand{\ee}{\end{eqnarray*}}
\newcommand{\bea}{\begin{eqalph}}
\newcommand{\eea}{\end{eqalph}}

\newcommand{\nw}[1]{[{#1}]}

\newcommand{\gO}{{\got O}}

\newcommand{\cO}{{\cal O}}
\newcommand{\cA}{{\cal A}}
\newcommand{\cG}{{\got g}}
\newcommand{\gd}{{\got d}}

\newcommand{\gE}{{\got E}}
\newcommand{\gQ}{{\got T}}

\newcommand{\cV}{{\cal V}}

\newcommand{\cQ}{{\cal T}}

\newcommand{\bb}{{\bf 1}}

\newcommand{\al}{\alpha}

\newcommand{\dl}{\delta}
\newcommand{\la}{\lambda}
\newcommand{\La}{\Lambda}
\newcommand{\f}{\phi}

\newcommand{\m}{\mu}

\newcommand{\g}{\gamma}
\newcommand{\G}{\Gamma}

\newcommand{\ve}{\varepsilon}
\newcommand{\th}{\theta}

\newcommand{\si}{\sigma}

\newcommand{\w}{\wedge}

\newcommand{\wt}{\widetilde}

\newcommand{\ol}{\overline}
\newcommand{\dr}{\partial}

\newcommand{\ar}{\op\longrightarrow}

\newcommand{\ot}{\otimes}

\newcounter{eqalph}
\newcounter{equationa}
\newcounter{example}
\newcounter{remark}
\newcounter{theorem}
\newcounter{proposition}
\newcounter{lemma}
\newcounter{corollary}
\newcounter{definition}

\def\thedefinition{\arabic{definition}}

\newenvironment{proof}{\noindent{\it Proof.}}{\hfill $\Box$
\medskip }
\newenvironment{rem}{\medskip \noindent{\bf Remark.
}}{ \medskip }

\newenvironment{prop}{\refstepcounter{definition} \medskip\noindent{\bf
Proposition \thedefinition.}\sl}{\medskip }
\newenvironment{lem}{\refstepcounter{definition} \medskip\noindent{\bf  Lemma
\thedefinition.}\sl }{\medskip }
\newenvironment{cor}{\refstepcounter{definition} \medskip\noindent{\bf 
Corollary \thedefinition.}\sl }{\medskip }

\newenvironment{eqalph}{\stepcounter{equation}
\setcounter{equationa}{\value{equation}}
\setcounter{equation}{0}

\begin{eqnarray}}{\end{eqnarray}\setcounter{equation}{\value{equationa}}}

\newcommand{\bs}{{\bf s}}

\newcommand{\mar}[1]{}

\begin{document}

{\parindent=0pt

{\large\bf Cohomology of the variational complex in 
BRST theory} 
\bigskip

{\sc G. Sardanashvily}

Department of Theoretical Physics, 
Moscow State University, 117234 Moscow, Russia

E-mail: sard@grav.phys.msu.su
\bigskip\bigskip

{\small

{\bf Abstract}. 
We show that cohomology of the variational complex in the field-antifield BRST
theory on an arbitrary manifold is equal to the de Rham cohomology of 
this manifold.
} }

\section{Introduction}

In the field-antifield BRST theory, the antibracket and the BRST operator 
are
defined by means of the variational operator (see, e.g., \cite{gom}). 
To introduce this variational operator in a rigorous algebraic way, one
can replace  
the calculus in functionals with the
calculus in jets of fields and antifields, and can construct the
variational complex 
\cite{barn,barn00,brandt97}. Furthermore, one has proved that the
variational complex in BRST theory on a
contractible manifold $\Bbb R^n$ is exact
\cite{brandt,brandt97,drag}. This means that the kernel of the
variational operator $\dl$ coincides with the image of the horizontal (or
total) differential $d_H$. Therefore, main objects in the 
field-antifield BRST theory on $\Bbb R^n$ are
defined modulo $d_H$-exact forms. Let us mention, e.g., the local BRST
cohomology.

Here, the variational complex in the field-antifield BRST theory on
an arbitrary smooth manifold $X$ is studied; that requires a (global)
differential geometric definition of ghosts, antifields, and their jets.
We show that
cohomology of this variational complex equals the de Rham
cohomology of $X$. In other words, the obstruction to the exactness of
the variational complex in BRST theory lies only in closed non-exact
forms on $X$. This fact 
enables one to generalize many constructions of the field-antifield
BRST theory on $\Bbb R^n$ to that on an arbitrary manifold $X$. 
In particular, global
descent equations on $X$ can be defined \cite{lmp}. 

For the sake of simplicity, we will consider the case of
even physical fields and even irreducible gauge
transformations with a finite number of generators. Then ghosts are
odd, and antifields are odd and even. For
instance, this is the case of the Yang--Mills theory. One says that
physical fields, ghosts and antifields constitute a physical basis.

We start from cohomology of the variational complex of even classical fields.
Cohomology of the variational complex in BRST theory is studied in a
similar way.

\section{The variational complex of classical fields}

In classical field theory, fields are represented by sections of a
smooth fibre bundle $Y\to X$. Put further dim$\,X=n$. 

\begin{rem}
Smooth manifolds throughout are 
real, finite-dimensional, Hausdorff, second-countable (i.e.,
paracompact), and connected. The standard
notation of jet formalism is utilized (see, e.g., \cite{book,book00}).
We follow the terminology of
\cite{bred,hir}, where a sheaf $S$ is a particular topological bundle and 
$\G(S)$ denotes the group of global sections of $S$. 
\end{rem}

The configuration space of Lagrangian formalism on a fibre bundle
$Y\to X$ is  
the infinite order jet space $J^\infty Y$ of 
$Y\to X$. It is defined as a projective limit
of the inverse system
\mar{t1}\beq
X\op\longleftarrow^\pi Y\op\longleftarrow^{\pi^1_0}\cdots 
J^{r-1}Y \op\longleftarrow^{\pi^r_{r-1}} J^rY\longleftarrow\cdots \label{t1}
\eeq
of finite order jet manifolds $J^rY$ of $Y\to X$, where $\pi^r_{r-1}$
are affine bundles. One can say that 
$J^\infty Y$ consists of the equivalence classes of sections of $Y\to X$
identified by their Taylor series at points of $X$. Endowed with the
projective limit topology, the ionfinite order jet space
$J^\infty Y$ is a paracompact Fr\'echet manifold \cite{tak2}.
A bundle coordinate atlas
$\{U_Y,(x^\la,y^i)\}$ of $Y\to X$ yields the manifold
coordinate atlas
\be
\{(\pi^\infty_0)^{-1}(U_Y), (x^\la, y^i_\La)\}, \qquad 0\leq|\La|,
\ee
 of $J^\infty
Y$, together with the transition functions  
\mar{55.21}\beq
{y'}^i_{\la+\La}=\frac{\dr x^\m}{\dr x'^\la}d_\m y'^i_\La, \label{55.21}
\eeq
where $\La=(\la_k\ldots\la_1)$, $\la+\La=(\la\la_k\ldots\la_1)$ are
multi-indices and
\be
d_\la = \dr_\la + \op\sum_{|\La|\geq 0} y^i_{\la+\La}\dr_i^\La
\ee
is the total derivative.

Let us introduce the differential calculus on $J^\infty Y$.
With the inverse system (\ref{t1}), one has
the direct system 
\be
\cO^*(X)\op\longrightarrow^{\pi^*} \cO^*_0 
\op\longrightarrow^{\pi^{1*}_0} \cO_1^*
\op\longrightarrow^{\pi^{2*}_1} \cdots 
 \cO_r^* \op\longrightarrow^{\pi^{r+1*}_r}\cdots 
\ee
of differential algebras $\cO^*_r$ of exterior forms on finite
order jet manifolds $J^rY$, where $\pi^{r*}_{r-1}$ are the pull-back
monomorphisms. The direct limit 
of this direct system is the 
differential algebra $\cO^*_\infty$  which consists 
of all exterior forms on  finite order jet manifolds modulo the pull-back
identification. In particular, $\cO^*_\infty$ is the ring of the
pull-back onto $J^\infty Y$ of smooth real functions on finite order
jet manifolds.

For short, we
agree to call elements of $\cO^*_\infty$ the
exterior forms on
$J^\infty Y$. Of course, these forms are of bounded jet order.
Restricted to a coordinate chart
$(\pi^\infty_0)^{-1}(U_Y)$ of $J^\infty Y$, they
can be written in a coordinate form, where horizontal forms
$\{dx^\la\}$ and contact 1-forms
$\{\th^i_\La=dy^i_\La -y^i_{\la+\La}dx^\la\}$, together with the constant
function $\bb$, constitute the set of generators of the $\cO^0_\infty$-algebra
$\cO^*_\infty$. 
There is the canonical splitting
\be
\cO^*_\infty =\op\oplus_{k,s}\cO^{k,s}_\infty, \qquad 0\leq k, \qquad
0\leq s\leq n,
\ee
of $\cO^*_\infty$ into $\cO^0_\infty$-modules $\cO^{k,s}_\infty$
of $k$-contact and $s$-horizontal forms, together with the corresponding
projections
\be
&& h_k:\cO^*_\infty\to \cO^{k,*}_\infty, \quad 0\leq k, \\
&& h^s:\cO^*_\infty\to \cO^{*,s}_\infty, \quad 0\leq s
\leq n.
\ee 
Accordingly, the
exterior differential on $\cO_\infty^*$ is
decomposed into the sum $d=d_H+d_V$ of horizontal and vertical
differentials 
\be
&& d_H\circ h_k=h_k\circ d\circ h_k, \qquad d_H(\f)=
dx^\la\w d_\la(\f), \\ 
&& d_V \circ h^s=h^s\circ d\circ h^s, \qquad
d_V(\f)=\th^i_\La \w \dr^\La_i\f, \qquad \f\in\cO^*_\infty.
\ee

Lagrangians,
Euler--Lagrange operators, and other objects of a familiar
Lagrangian field theory are elements of the
differential algebra
$\cO^*_\infty$. They can be introduced in an algebraic way by
constructing the variational complex of the algebra $\cO^*_\infty$.

The $\Bbb R$-module 
endomorphism 
\be
&& \tau=\op\sum_{k>0} \frac1k\ol\tau\circ h_k\circ h^n, \\ 
&&\ol\tau(\f)
=(-1)^{\nm\La}\th^i\w [d_\La(\dr^\La_i\rfloor\f)], \qquad 0\leq\nm\La,
\qquad \f\in \cO^{>0,n}_\infty,
\ee
of $\cO^*_\infty$ is defined (see, e.g., \cite{bau,book,tul}). 
It is a projector, i.e., $\tau\circ\tau=\tau$, and
obeys the relations
\mar{z1}\beq
\tau\circ d_H=0, \qquad \tau\circ d\circ \tau-\tau\circ d=0. \label{z1}
\eeq
Put $E_k=\tau(\cO^{k,n}_\infty)$.
The variational operator on $\cO^{*,n}_\infty$ is defined as the
morphism $\dl=\tau\circ d$. 
It is nilpotent, and has the property
\mar{am13}\beq
\dl\circ\tau-\tau\circ d=0. \label{am13}
\eeq
Since the operators $d_H$ and $\dl$ are nilpotent, and the relations
(\ref{z1}) hold, we have the complex
\mar{tams}\beq
 0\to\Bbb R\to \cO^0_\infty \ar^{d_H}\cO^{0,1}_\infty\ar^{d_H}\cdots  
\op\longrightarrow^{d_H} 
\cO^{0,n}_\infty  \op\longrightarrow^\dl E_1 
\op\longrightarrow^\dl 
E_2 \longrightarrow \cdots, \label{tams}
\eeq
called the variational complex. Elements of its term
$\cO^{0,n}_\infty$ 
are Lagrangians, while those of $E_1$ are Euler--Lagrange operators.
There are the well-known statements summarized usually as
the algebraic Poincar\'e lemma (see, e.g., \cite{olver,tul}). 

\begin{lem} \label{am12} \mar{am12}
If $Y$ is a contractible fibre bundle $\Bbb R^{n+m}\to\Bbb R^n$, the
variational complex (\ref{tams}) is exact.
\end{lem}

To obtain cohomology of the variational complex (\ref{tams}) in the
case of an arbitrary smooth fibre bundle $Y\to X$, let us enlarge the
differential algebra $\cO^*_\infty$ as follows \cite{arx,lmp}. 

Let $\gQ^*_r$ be the sheaf
of germs of  exterior forms on $J^\infty Y$ and $\G(\gQ^*_\infty)$
the differential algebra of its global 
sections. One can say that the algebra $\G(\gQ^*_\infty)$
consists of exterior forms on $J^\infty Y$ which coincide locally (i.e., around
each point of $J^\infty Y$) with the pull-back of exterior
forms on finite-order jet manifolds. 
In particular, $\G(\gQ^0_\infty)$ is the ring of real functions on
$J^\infty Y$ such that, given
$f\in
\G(\gQ^0_\infty)$ and any point $q\in J^\infty Y$, there exists a neighborhood
of $q$ where $f$ coincides with the pull-back of a smooth function on some
finite order jet manifold.  
There is the natural monomorphism
$\cO^*_\infty \to\G(\gQ^*_\infty)$. 

Note that, in comparison with $\cO^0_\infty$, the jet order of elements
of $\G(\gQ^0_\infty)$ need not be bounded. Therefore, the algebra
$\G(\gQ^0_\infty)$ has a limited physical application. We involve it because
the paracompact space
$J^\infty Y$ admits a partition of unity by elements of the ring
$\G(\gQ^0_\infty)$ \cite{tak2}. It follows that the sheaves of
$\G(\gQ^0_\infty)$-modules on
$J^\infty Y$ are fine and, consequently, acyclic. Therefore, the
abstract de Rham theorem \cite{hir} can be called into play in order to
obtain cohomology of the differential algebra $\G(\gQ^*_\infty)$.
Then one proves that the algebras $\cO^*_\infty$ and $\G(\gQ^*_\infty)$ have
the same cohomology.

Since $\tau$ and $\dl$ on $\cO^*_\infty$ are pointwise
operators, their direct limits are defined on the sheaf $\gQ^*_\infty$
and possess the properties (\ref{z1}) and (\ref{am13}). Then
we have the variational complex of sheaves
\mar{tams1}\beq
0\to\Bbb R\to \gQ^0_\infty \ar^{d_H}\gQ^{0,1}_\infty\ar^{d_H}\cdots  
\op\longrightarrow^{d_H} 
\gQ^{0,n}_\infty  \op\longrightarrow^\dl \gE_1 
\op\longrightarrow^\dl  \cdots  \label{tams1}
\eeq
and the corresponding complex of differential algebras of their global
sections 
\mar{tams'}\beq
0\to\Bbb R\to \G(\gQ^0_\infty) \ar^{d_H}\G(\gQ^{0,1}_\infty)\ar^{d_H}\cdots  
\op\longrightarrow^{d_H} 
\G(\gQ^{0,n}_\infty)  \op\longrightarrow^\dl \G(\gE_1) 
\op\longrightarrow^\dl \cdots\,.  \label{tams'}
\eeq
By virtue of the Lemma \ref{am12}, the variational complex
(\ref{tams1}) is exact. 
The sheaves $\gQ^{k,m}$ in this complex are sheaves of
$\G(\gQ^0_\infty)$-modules and, consequently, are fine. One can prove that 
the sheaves $\gE_k$, being projections $\tau(\gQ^{k,n}_\infty)$ of
sheaves of $\G(\gQ^0_\infty)$-modules, are also fine \cite{arx,lmp}.
Consequently, the variational complex 
(\ref{tams1}) is the fine resolution of the constant sheaf $\Bbb R$ on
$J^\infty Y$. Then we come to the following.

\begin{prop} \label{lmp05} \mar{lmp05}
Cohomology of the complex (\ref{tams'}) equals the de Rham cohomology
of the fibre bundle $Y$.
\end{prop}

\begin{proof}
By virtue of the above mentioned abstract de Rham theorem \cite{hir},
there is an isomorphism
between the cohomology of the complex (\ref{tams'}) 
and the cohomology $H^*(J^\infty Y,\Bbb R)$ of the paracompact space
$J^\infty Y$ with coefficients in the constant sheaf $\Bbb R$.
Since $Y$ is a strong deformation retract of $J^\infty Y$ \cite{ander},
the cohomology $H^*(J^\infty Y,\Bbb R)$ is isomorphic to the
cohomology $H^*(Y,\Bbb R)$ of $Y$ with
coefficients in the constant sheaf $\Bbb R$ \cite{bred} and,
consequently, to 
the de Rham cohomology $H^*(Y)$ of $Y$. 
\end{proof}

Proposition (\ref{lmp05}) recovers the results of \cite{ander80,tak2}, but
we also note the following.
Let us consider the de Rham complex of sheaves 
\beq
0\to \Bbb R\to
\gQ^0_\infty\op\longrightarrow^d\gQ^1_\infty\op\longrightarrow^d
\cdots
\label{lmp71}
 \eeq
on $J^\infty Y$ and the corresponding complex of differential algebras
\beq
0\to \Bbb R\to
\G(\gQ^0_\infty)\op\longrightarrow^d\G(\gQ^1_\infty)\op\longrightarrow^d
\cdots\,.
\label{5.13'}
\eeq
The complex (\ref{lmp71}) is exact due to
the Poincar\'e lemma, and is a fine resolution of the constant sheaf
$\Bbb R$ on $J^\infty Y$.  Then, similarly to Proposition \ref{lmp05},
we obtain that the de Rham cohomology 
of the differential algebra
$\G(\gQ^*_\infty)$  is isomorphic to that $H^*(Y)$ of the fibre bundle $Y$.
It follows that every closed form $\f\in \G(\gQ^*_\infty)$
is split into the sum
\beq
\si=\varphi +d\xi, \qquad \xi\in \G(\gQ^*_\infty), \label{tams2} 
\eeq
where $\varphi$ is a closed form on the fibre bundle $Y$.

The relation (\ref{am13}) for $\tau$ and
the relation $h_0d=d_Hh_0$ for $h_0$ define  a homomorphisms of the
de Rham complex (\ref{5.13'}) of the algebra $\G(\gQ^*_\infty)$ to its
variational 
complex (\ref{tams'}), and the corresponding homomorphism of their cohomology
groups is an isomorphism. Then, the splitting (\ref{tams2}) leads to
the following decompositions.

\begin{prop} \label{t41} \mar{t41}
Any $d_H$-closed form $\si\in\G(\cQ^{0,m})$, $m< n$, is represented by the sum
\mar{t60}\beq
\si=h_0\varphi+ d_H \xi, \qquad \xi\in \G(\gQ^{0,m-1}_\infty), \label{t60}
\eeq
where $\varphi$ is a closed $m$-form on $Y$.
Any $\dl$-closed form $\si\in\G(\gQ^{k,n})$, $k\geq 0$, is split into
\mar{t42}\ben
&& \si=h_0\varphi + d_H\xi, \qquad k=0, \qquad \xi\in \G(\cQ^{0,n-1}_\infty),
\label{t42a}\\ 
&& \si=\tau(\varphi) +\dl(\xi), \qquad k=1, \qquad \xi\in \G(\cQ^{0,n}_\infty),
\label{t42b}\\
&& \si=\tau(\varphi) +\dl(\xi), \qquad k>1, \qquad \xi\in \G(\gE_{k-1}),
\label{t42c}
\een
where $\varphi$ is a closed $n+k$-form on $Y$.
\end{prop}

Let us now return to the differential algebra $\cO^*_\infty$.
The following is proved \cite{arx,lmp}. 

\begin{prop} \label{am11} \mar{am11}
The differential algebra $\cO^*_\infty$  has the same $d$-, $d_H$- and
$\dl$-cohomology as $\G(\gQ^*_\infty)$.
\end{prop}

It follows that cohomology of the variational complex (\ref{tams}) of
the algebra $\cO^*_\infty$ is equal to the de Rham cohomology of the
fibre bundle $Y$. Furthermore, if $\si$ in decompositions (\ref{t60})
-- (\ref{t42c}) is an element of $\cO^*_\infty\subset
\G(\gQ^*_\infty)$, then $\xi$ is so.

In quantum field theory, all physical fields are linear or affine
quantities. Therefore, let
 $Y\to X$ is an affine bundle. Then $X$ is a strong deformation retract
of $Y$ and the de Rham cohomology of $Y$ is equal to
that of
$X$. In this case, cohomology of the variational complex (\ref{tams})
equals to the de Rham cohomology of the base manifold $X$.
Hence, every $d_H$-closed form $\f\in
\cO^{0,m<n}_\infty$ is split into the sum
\mar{aa3}\beq
\f=\varphi + d_H\xi, \qquad \xi\in \cO^{0,m-1}_\infty, \label{aa3}
\eeq
where $\varphi$ is a closed form on $X$. 
Any $\dl$-closed form $\si\in\cO^{0,n}$ is split into
\mar{aa'}\beq
\si=\varphi + d_H\xi, \qquad \xi\in \cO^{0,n-1}_\infty, \label{aa}
\eeq
where $\varphi$ is a non-exact $n$-form on $X$.

\section{Differential geometry of ghosts}

Different geometric models of odd ghosts have been suggested. For instance, 
a ghost field
in the Yang--Mills theory on a
principal bundle has been described as the Maurer--Cartan form on the gauge
group (see, e.g., \cite{bon,sch,thi}). This description however is
not extended to other gauge theories and to other odd elements of the
physical basis. 
We provide the following geometric model of odd fields on a smooth manifold
$X$ \cite{book00,sard00}.

Let 
$Y\to X$ be a vector bundle with an $m$-dimensional
typical fibre $V$ and $Y^*\to X$ the dual of $Y$. 
We consider the exterior bundle
\mar{z780}\beq
\w Y^*=\Bbb R\op\oplus_X(\op\oplus_{k=1}^m\op\w^k Y^*), \label{z780}
\eeq
whose typical fibre is the finitely generated Grassmann algebra $\w V^*$. 
Sections of the exterior bundle (\ref{z780}) are called graded
functions. Let $\cA_Y$ denote the sheaf of germs of 
graded functions on $X$.  The pair
$(X,\cA_Y)$ is a graded manifold with the body manifold
$X$ and the structure sheaf 
$\cA_Y$ \cite{bart,kost77}. We agree to call it a simple
graded manifold with the characteristic vector bundle $Y$. Note that
any graded manifold
$(X,\cA)$ is isomorphic to some simple graded manifold, but this
isomorphism fails to be canonical  \cite{bart,batch1}. 

Given a bundle atlas
$\{(U;x^\la,y^a)\}$ of
$Y$ with transition functions
$y'^a=\rho^a_b(x) y^b$, let
$\{c^a\}$ be the corresponding fibre bases for
$Y^*\to X$, together with the transition functions
$c'^a=\rho^a_b(x)c^b$. We will call $(x^\la, c^a)$ the local basis for the
simple graded manifold $(X,\cA_Y)$. With respect to this basis, graded
functions read 
\be
f=\op\sum_{k=0}^m \frac1{k!}f_{a_1\ldots
a_k}c^{a_1}\cdots c^{a_k}, 
\ee
where $f_{a_1\cdots
a_k}$ are local smooth real functions on $U$, and we omit the symbol of the
exterior product 
of coframes $c$. 

In BRST theory, the basis elements $c^i$ of a simple graded manifold
can describe odd ghosts. For instance, in the
Yang--Mills theory on a principal 
bundle $P\to X$ with the structure group $G$, the above bundle $Y$ is the Lie
algebra bundle $V_GP=VP/G$,
where $VP$ denotes the vertical tangent bundle of $P$. The typical
fibre of $V_GP$ is the right Lie algebra $\cG$ of the group $G$. If $X$ is a
compact manifold and $G$ is a semisimple matrix Lie group, the Sobolev
completion of the set of sections of $V_GP\to X$ is the Lie 
algebra of the gauge group. The typical fibre of the dual $V^*_GP$ of
$V_GP$ is the coalgebra $\cG^*$. Let $\{\ve_r\}$ be a basis for $\cG$,
$\{e_r\}$ the corresponding fibre bases for $V_GP$,  and $\{C^r\}$ the
dual coframes in $V^*_GP$. Elements $C^r$ of these coframes play the
role of ghosts in the BRST extension of the Yang--Mills theory. Indeed,
the canonical section 
$C=C^r\ot e_r$ of the  tensor product
$V_G^*P\ot V_GP$ is the above mentioned Maurer--Cartan form on the
gauge group which one regards as a ghost field. In the heuristic formulation
of BRST theory, $C$ plays the role of a generator of gauge
transformations with odd parameters, i.e., is the BRST operator.

Let $\gd\cA_Y$ be the sheaf of graded derivations
of the sheaf $\cA_Y$. Its sections are called graded vector fields
on the graded manifold $(X,\cA_Y)$ (or, simply, on $X$). Any graded
vector field 
$u$ on an open subset $U\subset X$ is a graded derivation of the
graded algebra $\G(U,\cA_Y)$ of local graded functions on $U$, i.e.,  
\be
 u(ff')=u(f)f'+(-1)^{\nw u\nw f}fu (f'), \qquad f,f'\in \G(U,\cA_Y),
\ee
where $[.]$ denotes the Grassmann parity.
The $\gd\cA_Y$ is a sheaf of Lie
superalgebras with respect to the bracket 
\be
[u,u']=uu' + (-1)^{\nw u\nw{u'}+1}u'u.
\ee

Graded vector fields on a simple graded manifold 
can be seen as sections of a vector bundle as follows.
Due to the canonical splitting
$VY\cong Y\times Y$, the vertical tangent bundle 
$VY\to Y$ of $Y\to X$ can be provided with the fibre bases $\{\dr/\dr
c^a\}$, dual of 
$\{c^a\}$. 
These are the fibre basis for $\pr_2VY\cong Y$.  Then
a graded vector field on a trivialization domain $U$ reads
\mar{hn14}\beq
u= u^\la\dr_\la + u^a\frac{\dr}{\dr c^a}, \label{hn14}
\eeq
where $u^\la, u^a$ are local graded functions \cite{bart,book00}.
It yields a derivation of $\G(U,\cA_Y)$ by the rule
\mar{cmp50'}\beq
u(f_{a\ldots b}c^a\cdots c^b)=u^\la\dr_\la(f_{a\ldots b})c^a\cdots c^b +u^d
f_{a\ldots b}\frac{\dr}{\dr c^d}\rfloor (c^a\cdots c^b). \label{cmp50'}
\eeq
This rule implies the corresponding
coordinate transformation law 
\be
u'^\la =u^\la, \qquad u'^a=\rho^a_ju^j +u^\la\dr_\la(\rho^a_j)c^j 
\ee
of graded vector fields. It follows that graded vector fields (\ref{hn14})
can be represented by
sections of the vector bundle
$\cV_Y\to X$ which is locally isomorphic to the vector bundle
\be
\cV_Y\mid_U\approx\w Y^*\op\ot_X(\pr_2VY\op\oplus_X TX)\mid_U,
\ee
and has the bundle coordinates $(x^\la_{a_1\ldots a_k},v^i_{b_1\ldots b_k})$,
$k=0,\ldots,m$, together with the transition functions
\be
&& x'^\la_{i_1\ldots i_k}=\rho^{-1}{}_{i_1}^{a_1}\cdots
\rho^{-1}{}_{i_k}^{a_k} x^\la_{a_1\ldots a_k}, \\
&& v'^i_{j_1\ldots j_k}=\rho^{-1}{}_{j_1}^{b_1}\cdots
\rho^{-1}{}_{j_k}^{b_k}\left[\rho^i_jv^j_{b_1\ldots b_k}+ \frac{k!}{(k-1)!} 
x^\la_{b_1\ldots b_{k-1}}\dr_\la\rho^i_{b_k}\right].
\ee

There is the exact sequence 
\be
0\to \w Y^*\op\ot_X\pr_2VY\to\cV_Y\to \w Y^*\op\ot_X TX\to 0
\ee
of vector bundles over $X$.
Its splitting 
\mar{cmp70}\beq
\wt\g:\dot x^\la\dr_\la \mapsto \dot x^\la(\dr_\la +\wt\g_\la^a\frac{\dr}{\dr
c^a}) \label{cmp70} 
\eeq
transforms every vector field $\tau$ on $X$
into the graded vector field 
\mar{ijmp10}\beq
\tau=\tau^\la\dr_\al\mapsto \nabla_\tau=\tau^\la(\dr_\la
+\wt\g_\la^a\frac{\dr}{\dr c^a}),
\label{ijmp10} 
\eeq
which is a graded derivation of the sheaf $\cA_Y$ satisfying the Leibniz rule
\be
\nabla_\tau(sf)=(\tau\rfloor ds)f +s\nabla_\tau(f), \quad f\in\G(U,\cA_Y),
\quad s\in C^\infty(X), 
\ee
for any open subset $U\subset X$. Therefore, one can think of the splitting
(\ref{cmp70}) as being a graded connection on the simple graded manifold
$(X,\cA_Y)$
\cite{book00}. It should be emphasized that this notion of a graded
connection differs from that 
of a connection on a graded fibre bundle in \cite{alm}.
In particular, every linear connection 
\be
\g=dx^\la\ot (\dr_\la +\g_\la{}^a{}_bv^b\dr_a) 
\ee
on the vector bundle $Y\to X$ yields the graded connection 
\mar{cmp73}\beq
\wt \g=dx^\la\ot (\dr_\la +\g_\la{}^a{}_bc^b\frac{\dr}{\dr c^a}). \label{cmp73}
\eeq

For instance, let $Y$ be the Lie algebra bundle $V_GP$ in the
Yang--Mills theory on a $G$-principal bundle $P$. Every principal
connection $A$ on $P\to X$ yields a linear connection 
\be
A=dx^\la\ot(\dr_\la - c^r_{pq}A^p_\la \xi^q e_r)
\ee
on $V_GP\to X$ \cite{book00} and, consequently, the graded connection
on ghosts 
\be
\wt A=dx^\la\ot(\dr_\la - c^r_{pq}A^p_\la C^q \frac{\dr}{C^r}),
\ee
where $c^r_{pq}$ are the structure constants of the Lie algebra $\cG$.

Let $\cV^*_Y\to  X$ be a
vector bundle which is the pointwise $\w Y^*$-dual of $\cV_Y$.
It is locally isomorphic to the vector bundle
\be
\cV^*_Y\mid_U\approx \w Y^*\op\ot_X(\pr_2VY^*\op\oplus_X T^*X)\mid_U.
\ee
With respect to the dual bases $\{dx^\la\}$ for $T^*X$ and
$\{dc^b\}$ for $\pr_2V^*Y=Y^*$, sections of the vector bundle $\cV^*_Y$
take the coordinate form 
\be
\f=\f_\la dx^\la + \f_adc^a,
\ee
together with transition functions 
\be
\f'_a=\rho^{-1}{}_a^b\f_b, \qquad \f'_\la=\f_\la
+\rho^{-1}{}_a^b\dr_\la(\rho^a_j)\f_bc^j.
\ee
They are treated as graded 1-forms on the graded manifold
$(X,\cA_Y)$. 

The sheaf $\gO^1\cA_Y$ of germs of sections of the vector bundle
$\cV^*_Y\to X$ is the dual of the sheaf $\gd\cA_Y$, where
the duality morphism is given by the interior product 
\be
u\rfloor \f=u^\la\f_\la + (-1)^{\nw{\f_a}}u^a\f_a. 
\ee

Graded $k$-forms $\f$ are defined as sections
of the graded exterior bundle $\op\w^k_X\cV^*_Y$ such that
\be
\f\w\si =(-1)^{\nm\f\nm\si +\nw\f\nw\si}\si\w \f,
\ee
where $|.|$ denotes the form degree.
The graded exterior differential
$d$ of graded functions is introduced in accordance with the condition 
$u\rfloor df=u(f)$
for an arbitrary graded vector field $u$, and  is
extended uniquely graded exterior forms by the rules
\be
d(\f\w\si)= (d\f)\w\si +(-1)^{\nm\f}\f\w(d\si), \qquad  d\circ d=0.
\ee
It takes the coordinate form
\be
d\f= dx^\la \w \dr_\la(\f) +dc^a\w \frac{\dr}{\dr c^a}(\f), 
\ee
where the left derivatives 
$\dr_\la$, $\dr/\dr c^a$ act on coefficients of graded exterior forms
by the rule 
(\ref{cmp50'}), and they are graded commutative with the forms $dx^\la$,
$dc^a$. 

With $d$, graded exterior forms constitute a graded differential algebra
$\cO^*\cA_Y$, where $\cO^0\cA_Y=\G(\cA_Y)$ is the graded commutative
ring of graded functions on $X$. There
is a monomorphism of differential algebras $\cO^*(X)\to \cO^*\cA_Y$. 
Let $\gQ^*\cA_Y$ denote the sheaf of germs of
graded exterior forms on $X$. Then $\cO^*\cA_Y=\G(\gQ^*\cA_Y)$. 

If the basis elements $c^a$ of the graded manifold $(X,\cA_Y)$ are
treated as ghosts of ghost number 1, graded exterior forms $\f\in \cO^*\cA_Y$
can also be provided with a ghost number by the rule
\be
{\rm gh}(dc^a)=1, \qquad {\rm gh}(dx^\la)=0.
\ee
Then the Grassmann parity $[\f]$ is equal to gh$(\f)\,{\rm mod}2$.
Ona also introduces the total ghost number gh$(\f) +|\f|$. 

\section{Jets of ghosts}

As was mentioned above, the antibracket and the BRST opreator in the
field-antifield BRST theory of \cite{barn,barn00,brandt97} are expressed
in terms of jets of ghosts. 
For example, the BRST transformation of gauge
potentials $a^m_\la$ in the Yang--Mills theory reads
\be
\bs a^r_\la= C^r_\la + c^r_{pq}a^p_\la C^q,
\ee
where $C^r_\la$ are jets of ghosts $C^r$ introduced
usually in a heuristic way. 
We will describe jets of odd fields as elements of a particular simple graded
manifold. 

Let $Y\to X$ be the characteristic vector bundle  of a simple graded
manifold $(X,\cA_Y)$. The $r$-order jet
manifold $J^rY$ of $Y$ is also a vector bundle over $X$. Let us
consider the simple graded manifold $(X,\cA_{J^rY})$ with the
characteristic vector bundle
$J^rY\to X$. Its local basis is $\{x^\la,c^a_\La\}$, $0\leq |\La|\leq r$,
together with the transition functions
\mar{+471}\beq
c'^a_{\la +\La}=d_\la(\rho^a_j c^j_\La), \label{+471}
\eeq 
where 
\be
d_\la=\dr_\la + \op\sum_{|\La|<r}c^a_{\la+\La}\frac{\dr}{\dr c^a_\La}
\ee
denotes the graded total 
derivative. In view
of the transition functions (\ref{+471}), one can think of $(X,\cA_{J^rY})$ as
being a graded $r$-order jet manifold of the graded manifold
$(X,\cA_Y)$. It should be emphasized that this
notion differs from that of a graded jet manifold of a graded
fibre bundle \cite{rup}.

Let $\cO^*\cA_{J^rY}$ be the differential algebra of graded exterior
forms on the simple graded manifold $(X,\cA_{J^rY})$.
Being a linear bundle morphism
of vector bundles over $X$, 
the affine bundle $\pi^r_{r-1}:J^rY \to J^{r-1}Y$ yields the
corresponding morphism of simple graded manifolds 
$(X,\cA_{J^rY})\to (X,\cA_{J^{r-1}Y})$ \cite{book00} and the
pull-back  monomorphism of differential algebras
$\cO^*\cA_{J^{r-1}Y}\to \cO^*\cA_{J^rY}$. With the inverse system
of jet manifolds (\ref{t1}),
we have the direct system of differential algebras 
\be
\cO^*\cA_Y\ar \cO^*\cA_{J^1Y}\ar\cdots 
\cO^*\cA_{J^rY}\ar^{\pi^{r+1*}_r}\cdots\,.
\ee
Its direct limit $\cO^*_\infty\cA_Y$
consists 
of graded exterior forms on graded jet manifolds $(X,\cA_{J^rY})$,
$0\leq r$, modulo the pull-back identification. 
It is a locally free
$C^\infty(X)$-algebra generated by the elements 
\be
(1,dx^\la, c^a_\La, \th^a_\La=dc^a_\La -c^a_{\la +\La}dx^\la), \qquad
0\leq |\La|. 
\ee
We have the corresponding decomposition of $\cO^*_\infty\cA_Y$ into
$\cO^0_\infty\cA_Y$-modules $\cO^{k,s}_\infty\cA_Y$ of $k$-contact and
$s$-horizontal graded forms.
Accordingly, the graded exterior differential $d$ on the algebra
$\cO^*_\infty\cA_Y$ is split into the sum $d=d_H+d_V$ of the graded horizontal
differential 
\be
d_H(\f)=dx^\la\w d_\la(\f), \qquad \f\in \cO^*\cA_\infty,
\ee
and the graded vertical differential $d_V$.

If the basis elements $c^a$ of the graded manifold $(X,\cA_Y)$ are
treated as ghosts of ghost number 1, jets of ghosts $c^a_\La$ and the
graded exterior forms $dc^a_\La$ are also
provided with ghost number 1.

\section{Even fields and antifields}

In order to describe odd and even elements of the physical basis of
BRST theory  on the
same footing, let us generalize the notion of a graded manifold to graded
commutative algebras generated  both by odd and even elements
\cite{man99}. 

Let $Y=Y_0\oplus Y_1$ be the Whitney sum of vector bundles $Y_0\to
X$ and $Y_1\to X$. We treat it as
a bundle of graded vector spaces with the typical fibre $V=V_0\oplus V_1$.
Let us consider the quotient of the tensor bundle   
\be
\ot Y^*=\op\oplus^\infty_{k=0} (\op\ot^k_X Y^*)
\ee
by the elements 
\be
y_0y'_0 - y'_0y_0,\quad y_1y'_1 + y'_1y_1, \quad y_0y_1 -
y_1y_0
\ee 
for all  $y_0,y'_0\in Y_{0x}^*$,
$y_1,y'_1\in Y_{1x}^*$, and  $x\in X$.
This is an infinite-dimensional vector bundle which we will denote by $\w Y^*$.
Global sections of $\w Y^*$ constitute a graded commutative algebra
$\cA_Y(X)$ which is the product over $C^\infty(X)$ of the commutative algebra
$\cA_0(X)$ of global sections of the symmetric bundle $\vee Y_0^*\to X$ and
the graded algebra $\cA_1(X)$ of global sections of the exterior bundle $\w
Y_1^*\to X$.  

Let $\cA_Y$, $\cA_0$ and $\cA_1$ be the sheaves of germs of
sections of the vector bundles $\w Y^*$, $\vee Y_0^*$ and $\w Y_1^*$,
respectively. For instance, the pair 
$(X,\cA_1)$ is a familiar simple graded manifold.  For the sake of brevity, we
therefore agree to call 
$(X,\cA_Y)$ the graded commutative manifold with the characteristic vector
bundle $Y$. Given
a coordinate chart
$(x^\la,y^i_0,y^a_1)$ of $Y$, the local basis for $(X,\cA_Y)$ is
$(x^\la,c^i_0, c^a_1)$, where
$\{c^i_0\}$ and $\{c^a_1\}$ are the fibre bases for the vector bundles $Y_0^*$
and
$Y_1^*$, respectively. Then a straightforward repetition of all the above
constructions for a simple graded manifold provides us with the
differential algebra 
$\cO^*\cA_\infty$ of graded commutative exterior forms on $X$. This is a
$C^\infty(X)$-algebra generated locally by the elements
$(1,c^i_{0\La},c^a_{1\La}, dx^\la, \th^i_{0\La}, \th^a_{1\La})$, 
$0\leq |\La|$.
Its $C^\infty(X)$-subalgebra $\cO^*\cA_{1\infty}$, generated locally by
the elements $(1,c^i_{1\La}, dx^\la, \th^i_{1\La})$,
is exactly the differential algebra $\cO^*_\infty\cA_{Y_1}$ on the
simple graded manifold  
$(X,\cA_1)$. The
$C^\infty(X)$-subalgebra
$\cO^*\cA_{0\infty}$ of $\cO^*\cA_\infty$, 
generated locally by the elements
$(1,c^i_{0\La}, dx^\la, \th^i_{0\La})$, 
$0\leq |\La|$, is isomorphic to the
polynomial subalgebra of the differential algebra $\cO^*_\infty$ of
exterior forms on the infinite 
order jet manifold $J^\infty Y_0$ of the vector bundle $Y_0\to X$. This
isomorphism is 
performed by the formal assignment $y^i_{0\La}\leftrightarrow c^i_{0\La}$
which is preserved by the transition functions (\ref{55.21}) and
(\ref{+471}). 

In the field-antifield BRST theory, the basis elements $c^i_{0\La}$ of 
the algebra $\cO^*\cA_\infty$ can characterize even elements of the
physical basis and their jets, while $c^a_{1\La}$ describe odd elements
of the physical basis and their jets.

It should be emphasized that, in the jet formulation of the field-antifield
BRST theory, antifields
can be introduced on the same footing as physical fields and ghosts. 
Let us denote physical fields and ghosts by the collective symbol $\Phi^A$.
Let $E$ be the characteristic vector bundle of the graded
commutative manifold generated by $\Phi^A$.
Treated as source
coefficients of BRST transformations, antifields $\Phi^*_A$ are represented by
elements of the graded commutative manifold whose structure vector
bundle is $\op\w^n T^*X\ot E^*$ (cf. the geometric treatment of
antifields in functional BRST formalism \cite{khud,witt}). 

In
particular, gauge potentials in the Yang--Mills theory are represented
by sections of the affine bundle $J^1P/G\to X$ modelled on the vector
bundle $TX\ot V^*_GP$. Their antifields are the basis
elements of the vector bundle  $\op\w^n T^*X\ot T^*X\ot V_GP$.
Accordingly, the antifields of ghosts in the Yang--Mills theory are the
basis elements of the vector bundle $\op\w^n T^*X\ot V^*_GP$.

\section{The variational complex in BRST theory}

The differential algebra $\cO^*\cA_\infty$ gives everything that
 one needs for a global formulation of the Lagrangian field-antifield
BRST theory 
in jet terms. In particular, let us consider the short variational complex 
\mar{cmp25}\beq
0\ar \Bbb R\ar \cO^0\cA_\infty\ar^{d_H}\cO^{0,1}\cA_\infty\ar^{d_H}\cdots
\ar^{d_H} \cO^{0,n}\cA_\infty\ar^\dl \im\dl \to 0,
\label{cmp25}
\eeq
where $\dl$ is given by the expression 
\be
\dl (L)= 
 (-1)^{|\La|}\th^a\w d_\La (\dr^\La_a L), \qquad L\in \cO^{0,n}\cA_\infty,
\ee
with respect to a physical basis $\{\zeta^a\}$.
The variational complex (\ref{cmp25}) provides the
algebraic approach to the antibracket technique, where one can think of
elements of $\cO^{0,n}\cA_\infty$ as being Lagrangians of fields,
ghosts and antifields.

To obtain cohomology of the variational complex (\ref{cmp25}), one can
follow exactly the procedure in Section 2. Let us consider the sheaf
$\gQ^*\cA_\infty$ of germs of graded commutative exterior forms
$\f\in\cO^*\cA_\infty$ and the differential algebra
$\G(\gQ^*\cA_\infty)$ of global sections of this sheaf. 
We have the short variational complex of sheaves
\mar{cmp20}\beq
0\ar \Bbb R\ar \gQ^0\cA_\infty\ar^{d_H}\gQ^{0,1}\cA_\infty\ar^{d_H}\cdots
\ar^{d_H} \gQ^{0,n}\cA_\infty\ar^\dl \im\dl \to 0.
\label{cmp20}
\eeq
There is the following
variant of the algebraic Poincar\'e lemma \cite{brandt,brandt97,drag}.

\begin{lem} \label{cmp21} \mar{cmp21}
The complex (\ref{cmp20}) is exact.  
\end{lem}

Since $\gQ^{0,*}\cA_\infty$ are sheaves of $C^\infty(X)$-modules, they are
fine and acyclic. Without studying the acyclicity of the sheaf
$\im\dl$, we can apply a minor modification of the abstract de Rham theorem
\cite{arx,tak2} to the complex
(\ref{cmp20}), and obtain the following.

\begin{prop}
Cohomology of the complex
\mar{cmp22}\beq
0\ar \Bbb R\ar
\G(\gQ^0\cA_\infty)\ar^{d_H}\G(\gQ^{0,1}\cA_\infty)\ar^{d_H}\cdots 
\ar^{d_H} \G(\gQ^{0,n}\cA_\infty)\ar^\dl \im\dl \to 0
\label{cmp22}
\eeq
is isomorphic to the de Rham cohomology of $X$.
\end{prop}

This cohomology isomorphism is performed by a monomorphism of
the de Rham complex of exterior forms on $X$ to the complex (\ref{cmp22});
that leads to the following. 

\begin{cor} \label{cmp26} \mar{cmp26}
Every
$d_H$-closed form $\f\in\G(\gQ^{0,m<n}\cA_\infty)$
is split into the sum
\mar{tt72}\beq
\f=\varphi + d_H\xi, \qquad \xi\in \G(\gQ^{0,m-1}\cA_\infty), \label{tt72}
\eeq
where $\varphi$ is a closed $m$-form on $X$. Every
$\dl$-closed form $\f\in\G(\gQ^{0,n}\cA_\infty)$ is
split into the sum
\mar{tt72'}\beq
\f=\varphi + d_H\xi, \qquad \xi\in \G(\gQ^{0,n-1}\cA_\infty), \label{tt72'}
\eeq
where $\varphi$ is a non-exact $n$-form on $X$. 
\end{cor}

Turn now to the short variational complex (\ref{cmp25}).
Its cohomology is equal to that of the complex (\ref{cmp22}). The proof
of this fact is a repetition of that of Proposition \ref{am11} where
exterior forms on $J^\infty Y$ are 
replaced with graded commutative forms on $X$ and, accordingly, Lemma
\ref{cmp21} and Corollary \ref{cmp26} are quoted. It follows that a graded
commutative form exterior $\xi$ in the expressions (\ref{tt72}) and
(\ref{tt72'}) 
belongs to the algebra $\cO^*\cA_\infty$ whenever $\f$ does.

We also mention the important case of a BRST theory where Lagrangians
are independent on coordinates $x^\la$. 
Let us consider the subsheaf $\ol\gQ^*\cA_\infty$ of the sheaf
$\gQ^*\cA_\infty$ which consists of germs of $x$-independent graded
commutative exterior forms. Then we have the subcomplex  
\mar{cmp27}\beq
0\ar \Bbb R\ar
\ol\gQ^0\cA_\infty\ar^{d_H}\ol\gQ^{0,1}\cA_\infty\ar^{d_H}\cdots
\ar^{d_H} \ol\gQ^{0,n}\cA_\infty\ar^\dl \im\dl \to 0
\label{cmp27}
\eeq
of the complex (\ref{cmp20}) and the corresponding subcomplex 
\mar{cmp28}\beq
0\ar \Bbb R\ar
\G(\ol\gQ^0\cA_\infty)\ar^{d_H}\G(\ol\gQ^{0,1}\cA_\infty)\ar^{d_H}\cdots
\ar^{d_H} \G(\ol\gQ^{0,n}\cA_\infty)\ar^\dl \im\dl \to 0
\label{cmp28}
\eeq
of the complex (\ref{cmp22}) which consists of $x$-independent graded
commutative exterior forms. It is readily observed that these forms are
of bounded jet order and 
$\G(\ol\gQ^{0,*}\cA_\infty)\subset \cO^{0,*}\cA_\infty$, i.e., the
complex (\ref{cmp28}) is also a subcomplex of the short variational
complex (\ref{cmp25}).

The key point is that the complex of sheaves (\ref{cmp27}) fails
to be exact. 
The obstruction to its exactness at the
term
$\ol\gQ^{0,k}_\infty$ is provided by the germs of $k$-forms on $X$ with
constant coefficients \cite{barn00}. 
Let us denote the sheaf of such germs
by
$S^k_X$.  
We have the short exact sequences of sheaves 
\be
&& 0\to \im d_H \to \Ker d_H \to S^k_X \to 0, \qquad 0<k< n,\\
&& 0\to \im d_H \to \Ker \dl \to S^n_X \to 0
\ee
and the corresponding sequences of modules of their global sections
\be
&& 0\to \G(\im d_H) \to \G(\Ker d_H) \to \G(S^k_X) \to 0, \qquad 0<k<
n,\\
&& 0\to \G(\im d_H) \to \G(\Ker \dl) \to \G(S^n_X) \to 0,
\ee
which are exact because $S^{k<n}_X$ and $S^n_X$ are subsheaves of $\Bbb
R$-modules of the sheaves
$\Ker d_H$ and $\Ker\dl$, respectively. Therefore, the $k$th cohomology
group of the  complex (\ref{cmp28})
is isomorphic to the $\Bbb R$-module
$\G(S^k_X)$ of global constant
$k$-forms, $0<k\leq n$, on the manifold $X$. 
Thus, any $d_H$-closed graded commutative $k$-form, $0< k<
n$, and any $\dl$-closed graded commutative $n$-form $\f$ 
are split into the sum
$\f=\varphi + d_H\xi$ where $\varphi\in \G(S^k_X)$ and $\xi\in \G(
\gQ^{0,k-1}\cA_\infty)$.

Thus, we observe that the obstruction to the exactness of the
variational complex in the field-antifield BRST theory on an arbitrary
manifold $X$ lies only in exterior forms on $X$. In particular, it
follows that the
topological ambiguity of a proper solution of the master equation in
the Lagrangian BRST theory
reduces to exterior forms on $X$.

\end{document}